\documentclass[]{aa}
\usepackage{graphicx}
\usepackage{color}
\usepackage{natbib}
\begin{document}
\title{The rotation - Lithium depletion correlation in the $\beta$ Pictoris association and  LDB age determination}
\author{S.\,Messina\inst{1},  A.C.\,Lanzafame\inst{2,1},  G.A. Feiden\inst{3}, M.\,Millward\inst{4}, S.\,Desidera\inst{5}, A.\,Buccino\inst{6}, I.\,Curtis\inst{7}, E.\,Jofr\'e\inst{8,9}, P.\,Kehusmaa\inst{10}, B.J.\,Medhi\inst{11}, B.\,Monard\inst{12}, R.\,Petrucci\inst{8,9}}
\offprints{Sergio Messina}
\institute{INAF-Catania Astrophysical Observatory, via S.Sofia, 78 I-95123 Catania, Italy \\
\email{sergio.messina@oact.inaf.it}
\and   
Universit\'a di Catania, Dipartimento di Fisica e Astronomia, Sezione Astrofisica, via S. Sofia 78, I-95123 Catania, Italy\\
\email{a.lanzafame@unict.it}
\and
Department of Physics \& Astronomy, Uppsala University, Box 516, SE-751 20, Uppsala, Sweden\\
\email{gregory.feiden@physics.uu.se}
\and
York Creek Observatory, Georgetown, Tasmania, Australia \\
\email{mervyn.millward@yorkcreek.net}
\and  
INAF-Osservatorio Astronomico di Padova, Vicolo dell'Osservatorio 5, I-35122 Padova, Italy\\
\email{silvano.desidera@oapd.inaf.it}
\and
Instituto de Astronom\'ia y F\'isica del Espacio (IAFE-CONICET), Buenos Aires,
Argentina\\
\email{abuccino@iafe.uba.ar}
\and
 IAU Minor Planet Center code D79, 2 Yandra Street, Vale Park,
South Australia 5081
Australia \\
\email{Ivan.Curtis@keyworks.com.au}
\and
Observatorio Astron\'omico de C\'ordoba, Laprida 854, X5000BGR, C\'ordoba, Argentina\\
\email{jofre.emiliano@gmail.com; romina@oac.unc.edu.ar }
\and
Consejo Nacional de Investigaciones Cient\'ificas y T\'ecnicas (CONICET),  Argentina
\and
Harlingten Atacama Observatory, San Pedro de Atacama, Chile.\\
\email{petri@kehusmaa-astro.com}
\and
Aryabhatta Research Institute of Observational Sciences, Manora Peak, Nainital 263129, India\\
\email{biman@aries.res.in}
\and
Klein Karoo Observatory, Western Cape, South Africa\\
\email{bmonard@mweb.co.za}
}

\date{}
\titlerunning{Rotation - Lithium correlation in the  $\beta$ Pic association}
\authorrunning{S.\,Messina et al.}
\abstract {There is evidence in the 125-Myr Pleiades cluster, and more recently in the 5-Myr NGC\,2264 cluster, that rotation plays a key role in the Lithium (Li) depletion processes among low-mass stars. Fast rotators appear to be less Li-depleted than equal-mass slow rotators.}
{We intend to explore the existence of a Li depletion - rotation connection among the $\beta$ Pictoris members at an age of about 24 Myr, and to use such correlation  either to confirm or to improve \rm  the age estimate based on the Lithium Depletion Boundary (LDB) modeling.} {We have photometrically monitored all the known members of the  $\beta$ Pictoris association with at least  one  Lithium equivalent width (Li EW) measurement from the literature.} {We measured  the rotation periods of 30 members for the first time and retrieved from the literature the rotation periods for other 36 members, building a catalogue of 66 members with measured rotation period and Li EW.} {We find that in the 0.3 $<$ M $<$ 0.8 M$_\odot$ range, there is a strong correlation between rotation and Li EW. For higher mass stars, no significant correlation is found. For very low mass stars in the Li depletion onset, at about 0.1  M$_\odot$, data are too few to infer a significant correlation.  The observed Li EWs are compared with those predicted by the  Dartmouth stellar evolutionary models that incorporate the effects of magnetic fields.  After decorrelating the Li EW from the rotation period, we find that the hot side of the LDB is fitted well by Li EW values corresponding to an age  of 25$\pm$3 Myr in good agreement with independent estimates from the literature.}
\maketitle
\rm

\section{Introduction}
In recent years much attention has been payed to the 24$\pm$3\,Myr \citep{Bell15} young \object{$\beta$ Pictoris} stellar association. Several studies have allowed 
		to increase significantly the number of confirmed members and to discover many more candidate members. Just to mention the most recent works, we refer the readers to \citet{Lepine09}; \citet{Kiss11}; \citet{Schlieder10, Schlieder12}; \citet{Shkolnik12}; \citet{Malo13, Malo14a, Malo14b}.  The first comprehensive search for the rotation periods of the low-mass members of  $\beta$ Pictoris was carried out by \citet{Messina10, Messina11} who retrieved a total of 38 low-mass members (i.e., spectral types from late F to M) from the earlier compilations of \citet{Zuckerman04}, \citet{Torres06}, and \citet{Kiss11}. Their study provided the rotation periods of 33 out of 38 members.\\
In the light of the mentioned studies, we have again explored up to the most recent literature and, at this time, we could finally
compile a new list of  117 stars \rm among members and candidate members with spectral types later than about F3V. Then, we have started a new rotational study on this enlarged sample. To get the photometric rotation periods of our targets, we used our own observations, archive data, and also we made use of periods from the literature.\\
As result of our photometric investigation, we obtained the rotation periods of   112 out of 117 stars. \rm Specifically,
we measured for the first time the rotation periods of   56 stars. \rm For another 27 stars, we could confirm with our analysis of
new or archived data the values reported in the literature. For 29 stars we adopted the literature values. For the remaining 5 stars, our periodogram analysis did not provide the rotation period. The results of this investigation are presented in  the catalogue of photometric rotation period of the \object{$\beta$ Pictoris} association members \rm (Messina et al. 2016; Paper I) where we describe the photometric observations newly obtained, their reduction and analysis, and a detailed discussion of our results obtained for each individual star.\\
In this paper (Paper II), we focus on a sub-sample consisting of 66 members for which we know the rotation period and have one measurement at least (from the literature)  of the Lithium equivalent width (EW). This sub-sample is to date the largest of  any known young loose association. We will make use of it to investigate the  correlation between rotation and Li depletion \rm and to  compare with earlier results \rm the age of the $\beta$ Pictoris association obtained by the modeling of the Lithium Depletion Boundary (LDB),  after decorrelating the Li EW from rotation. \rm

\section{Sample description}
The sample under analysis consists of 66 members  with spectral 
type later than F3V with one measurement, at least, of the Li EW (see  Table 1). This sample, which has  significantly
increased with respect to earlier studies, makes a new investigation on the age
of the $\beta$ Pictoris association with the LDB modeling method necessary. In fact, earlier studies
have made use of a  smaller number of association members. \citet{Mentuch08} made use of a sample of 23 members;
\citet{Macdonald10} used a sample of only 10 members. \citet{Binks14} and 
 \citet{Malo14b} used, respectively, about 40 and 34 members. \\
Moreover,  we also now know the rotation periods of all these 66 stars, which put us in the position to get more accurate results
and to deeply explore the mechanism of Li depletion. More specifically, we measured for the first time the rotation 
periods of 30 out of 66  members. These new measurements will be presented in the mentioned
catalogue containing  the photometric rotation periods of all 112 members/candidate members of the association. 
For another 16 members, the rotation periods were taken from
\citet{Messina10, Messina11}. The remaining 20 rotation periods were taken from different literature sources.\\
This sample represents, to date, one of the three  largest
samples ever used to investigate the  correlation between rotation period and Li depletion. \rm The other samples consist of members of the \object{Pleiades} and of the \object{M34} stellar open clusters investigated by \citet{Gondoin14} using the rotation periods from \citet{Hartman10}, and of members of the \object{NGC\,2264} open cluster whose results on the Lithium-rotation connection are presented by \citet{Bouvier16}.
The earlier larger sample of Pleiades members analysed by \citet{Soderblom93} made use of the projected rotational velocity
($v\sin{i}$), which is less accurate owing to the uncertainty arising from the unknown $\sin{i}$.

\section{Li distribution}

 \begin{figure*}
\begin{minipage}{20cm}
\includegraphics[scale = 0.6, trim = 0 0 0 0, clip, angle=90]{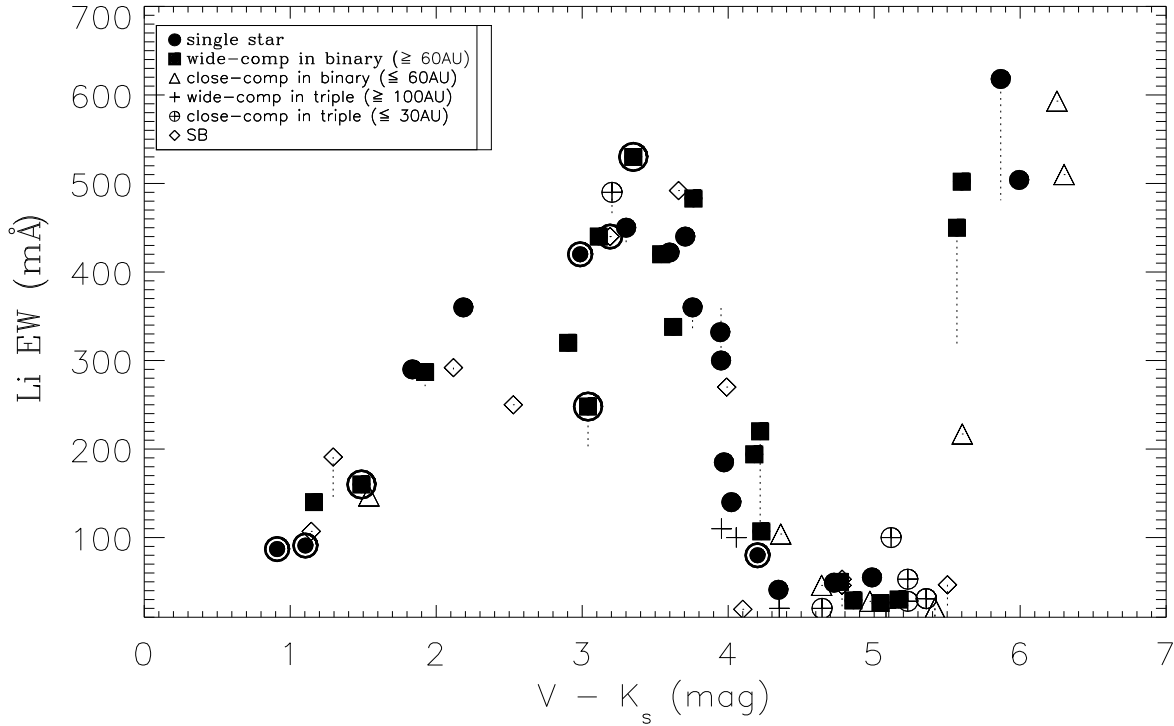}
\end{minipage}
\caption{\label{Li_distri} Distribution of Li EW versus V$-$K$_s$ color for confirmed members of the $\beta$ Pictoris association.
The meaning of symbols is given in the legend. Dotted vertical lines indicate the range of values for stars with multiple Li EW measurements. All circled symbols indicate stars hosting debris discs.}
\end{figure*}

In Fig.\,\ref{Li_distri}, we plot the distribution of Li EW for all 66 members versus the V$-$K$_s$ color. This color index is measured using the K$_s$ magnitude from the 2MASS project \citep{Cutri03} and the brightest (presumably unspotted) V magnitude from the long photometric time series available in the All Sky Automated Survey (ASAS) archive \citep{Pojmanski02}. This choice allows us to reduce the impact of photometric variability on the V-band magnitude. Among different color indexes, the V$-$K$_s$  turned out to be the most 
homogeneous and that with the largest range of values, allowing an accurate representation of different masses. We applied no color correction for interstellar absorption since   the members under study have distances in the range from 10\,pc to 80\,pc, \rm therefore the interstellar reddening can be considered negligible. The average uncertainty on this color is $\sigma_{\rm V-K_s}$ = 0.03\,mag. There are only 8 stars that belong to close binaries that are not resolved neither in V nor in K$_s$ magnitudes and consist of components differing by no more than 2 spectral sub-classes. For these stars the measured V$-$K$_s$ for the primary component, the one with measurement of Li EW, resulted to be redder than the expected value for the spectral type. In these cases we adopted the V$-$K$_s$ values from \citet{Pecaut13}, and corresponding to the spectral type of the primary component. \rm The Li EW measurements are retrieved from \citet{Mentuch08}, \citet{dasilva09}, \citet{Kiss11}, \citet{Binks14}, \citet{Moor13}, \citet{Malo14b}, \citet{Reid02}, \citet{Torres06}, and \citet{Desidera15}. The average uncertainty associated to these measurements is generally not larger than $\sim$40\,m\AA. The error bars associated to the measurements (in both axes) are equal to the size of the symbols plotted in Fig.\,\ref{Li_distri}. In a few cases more than one measurement is available. The range of values of the Li EW in these cases is plotted with a dotted line. Almost all measurements   (for a total of 20 members) \rm in the color range 4.5 $<$ V$-$K$_s$ $<$ 5.3\,mag are upper limits.
Different symbols are used to distinguish the different nature of the considered members (see the legend in Fig.\,\ref{Li_distri}).
Detailed information on each member and the criterion to distinguish close from wide components is given in Messina et al. (Paper III, in prep.).\\
In Fig.\,\ref{Li_distri}, we note three important features. First, we note a very well defined LDB that will be used for an estimate of the age. 
Second, in the color range  3.2 $<$ V$-$K$_s$ $<$ 4.5\,mag, we note, even by just a visual inspection, a scatter in the Li EW significantly larger than in the range of bluer members. We will investigate the role played by the rotation in producing this scattered depletion. 
Finally, we note that in the range 2.5 $<$V$-$K$_s$ $<$ 3.0\,mag, we have only three stars that all have Li EWs significantly lower than the average. This color range corresponds to K2--K3 spectral types and is reminiscent of the Li dip observed in more evolved main-sequence stars \citep[see, e.g.,][]{Balachandran95}.
One of these stars is \object{HIP\,84586}, a SB2 spectroscopic binary whose tidal locking between the two components may have altered the internal mixing and, consequently, enhanced the Li depletion rate with respect to single stars. The other two stars are \object{TYC\,6878\,0195\,1} and \object{HIP\,11437}A, both components of wide binaries that are expected to have evolved, rotationally, as single stars. We only note that \object{HIP\,11437}A hosts a debris disc.

\section{The rotation - Li depletion correlation}
We first   analyse the possible correlation between rotation and Li EW to investigate which role the rotation may play
in producing the observed Li EW dispersion among stars with similar masses. \\ \rm
This information will be used for a more accurate modeling of the LBD to infer the association age.

\begin{figure*}
\begin{minipage}{20cm}
\includegraphics[scale = 0.6, trim = 0 0 0 0, clip, angle=90]{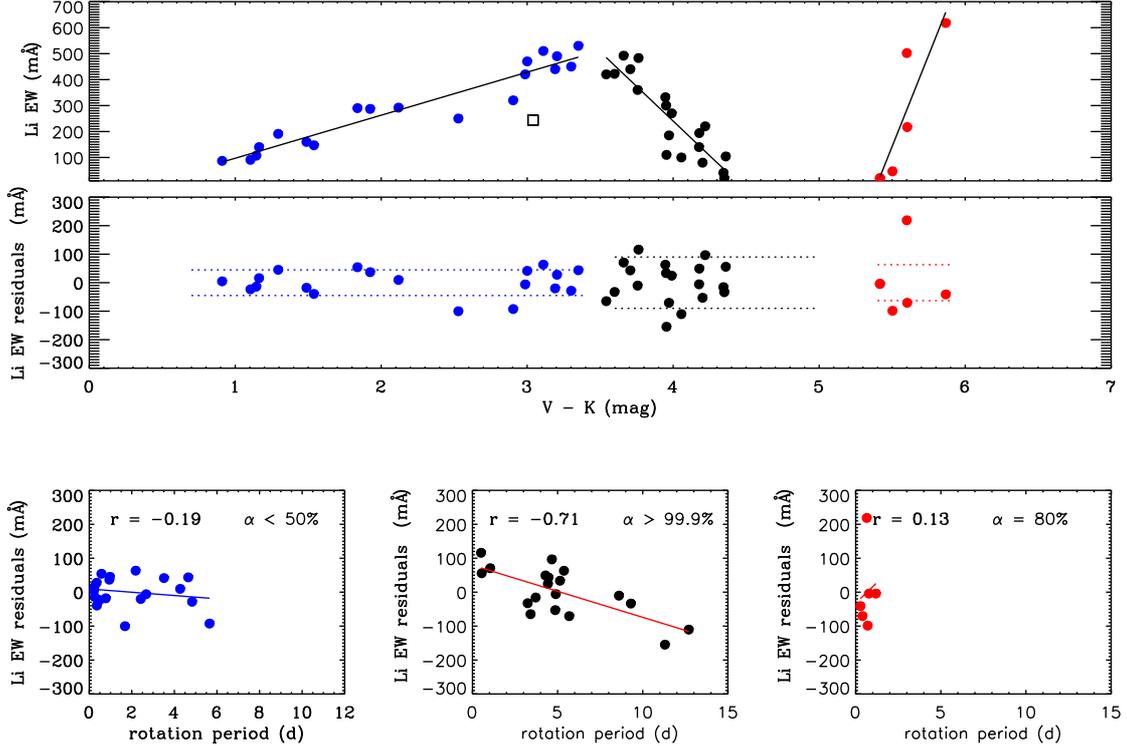}
\end{minipage}
\caption{\label{li_fit} \it  Top panel: \rm distribution of Li EW versus V$-$K$_s$ color in three color ranges: 0.5 $<$V$-$K$_s$ $<$ 3.4\,mag  (19 stars)\rm,  3.4 $<$V$-$K$_s$ $<$ 4.5\,mag (21 stars)\rm, and 5.4 $<$V$-$K$_s$ $<$ 5.9\,mag  (6 stars) \rm of $\beta$ Pictoris members with known rotation period and with overplotted (solid lines) the linear fits. \it  Middle panel: \rm residuals from the fit  with horizontal dotted lines indicating the 1$\sigma$ dispersion. \it Bottom panels: \rm distribution of Li EW residuals versus rotation period in the three color ranges. Solid lines are linear fits, r is the Pearson linear correlation coefficient and $\alpha$ the significance level.}
\end{figure*}

We analyse separately the following three sub-groups: blue stars  with 0.5 $\le$ V$-$K$_s$ $<$ 3.4\,mag  totaling 19 members, \rm red stars with 3.4 $\le$ V$-$K$_s$ $\le$ 4.5\,mag  totaling 21 members, \rm and very red stars with 5.4 $<$V$-$K$_s$ $<$ 5.9\,mag  totaling 6 members \rm that correspond to the onset of the Li depletion. Stars with V$-$K$_s$ $>$ 5.9\,mag have not undergone any Li depletion. In the blue stars sample, we observe a trend of increasing Li EW with increasing color (decreasing mass); in the red stars sample, we observe the opposite trend;  in the very red star sample we observe again  a trend of increasing Li EW with increasing color. As it is shown in the top panel of Fig.\,\ref{li_fit}, linear fits are enough for our purposes, i.e. approximating the mass dependence  to outline the correlation with the rotation period. In the middle panel we plot the residuals with respect to the linear fits. Only one star (\object{HIP\,11437}A), represented by a square and significantly deviating from the trend, was excluded from the fit. Positive residuals indicate stars that are more Li rich than the average, whereas negative residuals indicate stars that are more Li-depleted than the average.
In the case of blue stars, we find a peak-to-peak amplitude of the dispersion $\sigma$ = 90\,m\AA\,\, of the residuals from the fit, whereas in the case of red stars the peak-to-peak amplitude of the dispersion is about twice as large, $\sigma$ = 180\,m\AA. Finally, in the case of very red stars we find  $\sigma$ = 125\,m\AA.\,The dotted lines represent peak-to-peak amplitudes of dispersions in the three color ranges. Now, to probe any dependence of the residual dispersion on the rotation period, we computed the Pearson linear correlation coefficients (r) and significance levels\footnote{The significance
level $\alpha$ represents the probability of observing a value
of the correlation coefficient larger than $r$ for a random sample
having the same number of observations and degrees of freedom
\citep{Bevington69}} ($\alpha$) between the residuals
and the rotation period as well as the Spearman's rank correlation \citep{Spearman04}. In the case of blue stars we find $r$  = $-0.19$ with a significance level $\alpha <$ 50\%, i.e. no dependence
exists of the residuals on rotation (from Spearman's rank correlation we found $\rho$ = $-$0.04  with very small significance p-value  = 0.77). On the contrary, for red stars we found  $r$ = $-0.71$ with very high significance level $\alpha >$ 99.9\% (similarly, from rank correlation we found  $\rho$ = $-$0.56  with high significance p-value   = 0.01). Finally, for very red stars, we find $r$  = $+0.13$ with a significance level $\alpha \sim$ 80\% (from Spearman's rank correlation we found $\rho$ = 0.29  with very small significance p-value  = 0.58). 
Therefore, since in this regime of very red stars we have no enough data, we can not asses the existence or not of a rotation - Li depletion connection.\\
In the case of red stars (3.2 $\le$ V$-$K$_s$ $\le$ 4.5\,mag) we can state that  the Li depletion is significantly correlated to rotation \rm that seems to maintain fast rotators less
depleted than slow rotators according to Eq.\,(1):\\
\begin{equation}
\delta EW(Li) = 101\pm22 - 17.7\pm3.4\times P
\end{equation}
where the Li EW is in m\AA\,\, and the rotation period P in days. The average uncertainty on the rotation period ($\Delta$P = 0.7\%) is negligible with respect to that on the Li EW.
After  decorrelating  the Li EW from rotation of red stars, \rm the dispersion turns to be reduced by about a factor of 2, $\sigma$ = 50 \,m\AA\,\,
and becomes to be comparable to that measured among blue stars,  where no dependence on rotation was observed, and among very red stars.
This residual dispersion is larger than the uncertainty associated to the Li EW measurements.
Therefore, apart from mass and rotation, other causes must play a role in producing the observed dispersion. The intrinsic variability of the Li EW is certainly one cause. In fact, we know that spotted stars, as all our targets are, exhibit a rotational modulation of the Li EW. Specifically, the EW is larger at the rotation phases when the spots are best in view, and, consequently, the average surface temperature is lower.  
 In our sample 26 stars have two EW measurements. We find that the average difference between these multiple measurements is $<$$\Delta$EW$>$ = 26\,m\AA\,\,.  We note only two extreme cases: \object{TX Psa} and \object{2MASS\,J05082729-2101444} both very red stars with  $\Delta$EW $>$ 100\,m\AA. \rm\\
 Generally, the rotational Li  variability in young stars is not larger than about 5\% \citep[see Appendix in][]{Bouvier16}. \rm This corresponds in our case to a range from $\sim$5 to 25\,m\AA\,\, in absolute values, which is as the same order of the intrinsic variability we found.

\begin{figure*}
\begin{minipage}{22cm}
\includegraphics[scale = 0.3, trim = 0 0 0 0, clip, angle=90]{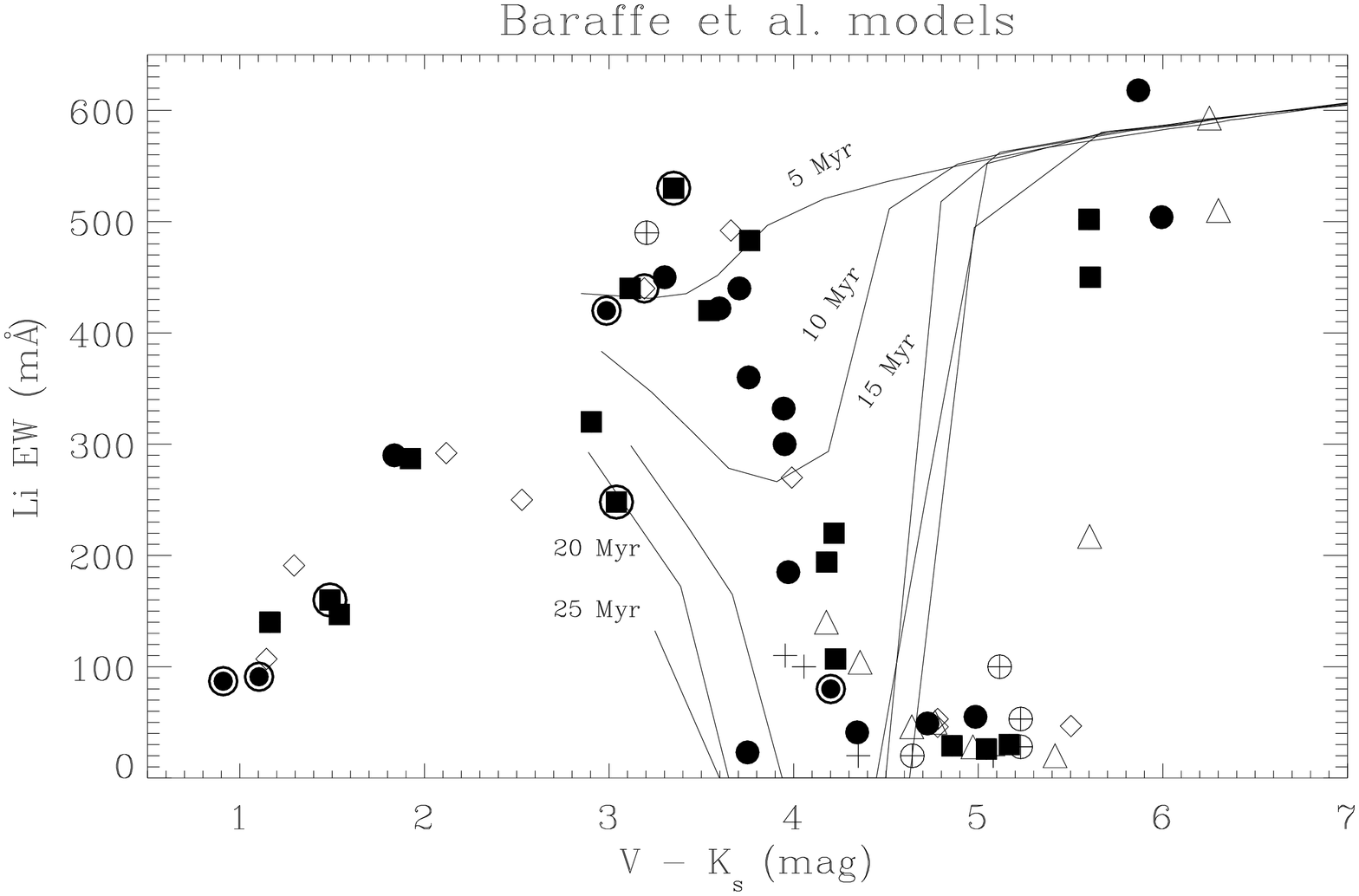}
\includegraphics[scale = 0.3, trim = 0 0 0 0, clip, angle=90]{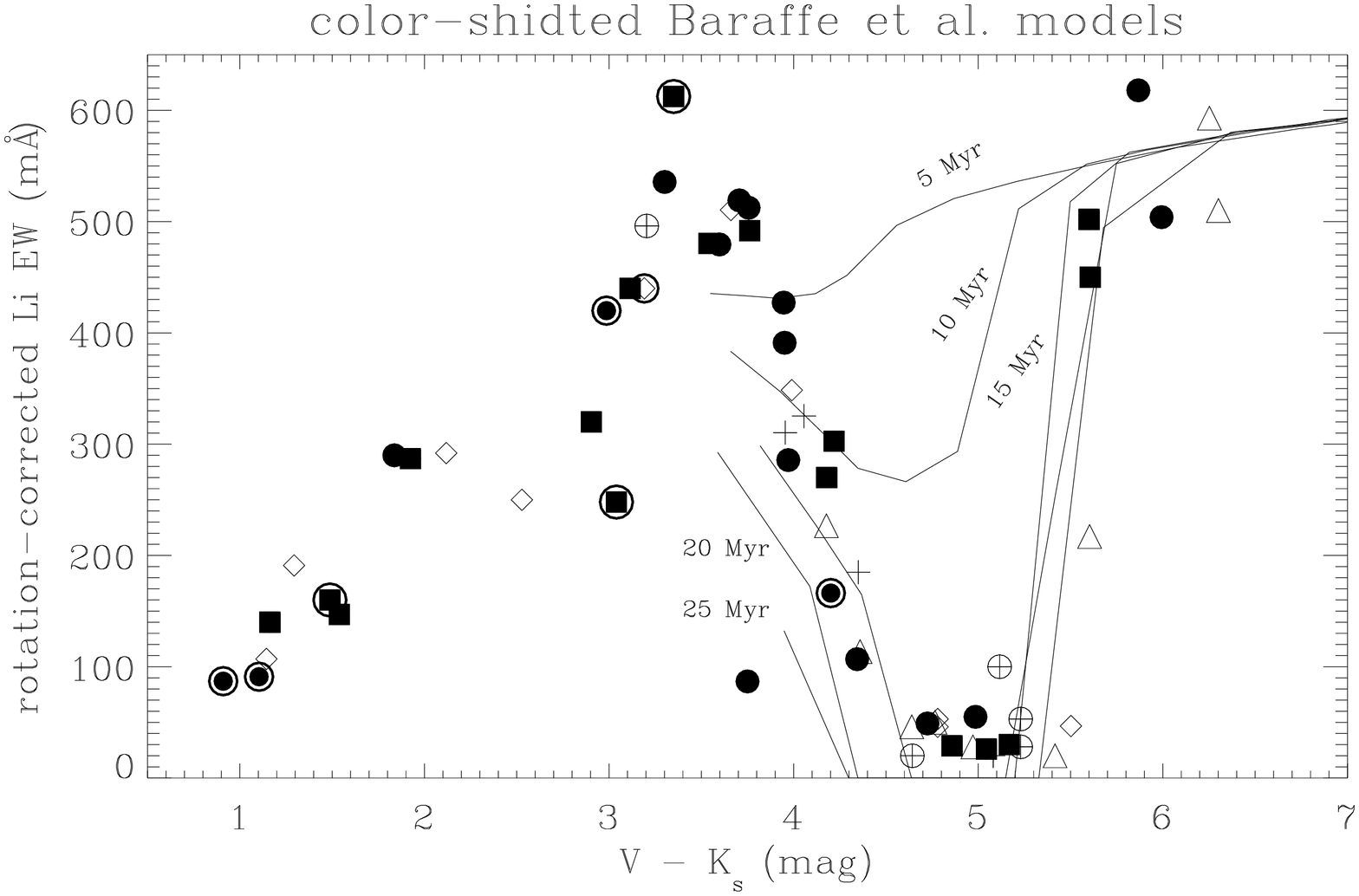}
\end{minipage}
\caption{\label{LDB} \it Left panel: \rm  distribution of Li EWs versus V$-$K$_s$ color for $\beta$ Pic members. Solid lines represent the Li EWs predicted by the models of \citet{Baraffe15} for ages in the range from 5 to 25 Myr. \it Right panel: \rm same as in the left panel, but with models arbitrarily shifted by 0.7\,mag towards redder colors to match with the observations.}
\end{figure*}

\subsection{Observed versus predicted LBD}
It is interesting to make a comparison between the measured distribution of Li EWs and the predictions of evolutionary models. In the following, we first consider the evolutionary models of \citet{Baraffe15}  for solar abundances. Solar abundances have been measured in a few members, like \object{$\beta$ Pic} and \object{PZ Tel}, and generally they were adopted for the whole association \citep[see, e.g.,][]{Mentuch08}.  The surface Lithium abundance is first derived from the ratio of surface Lithium abundance to initial abundance provided with the models using an initial abundance of Li$_0$ = 1$\times$10$^{-9}$.  \rm Then, to make the comparison with the observations, we transformed the model Li abundance into Li EW. For this purpose, we have used the curves of growth from \citet{Zapatero02}. They are valid in the effective temperature range 2600 $<$ T$_{\rm eff}$ $<$ 4100\,K and for 1.0 $<$ A(Li) $<$ 3.4.

In the left panel of Fig.\,\ref{LDB}, we plot the measured Li EW and compare this distribution with the  Li EW derived \rm from  \citet{Baraffe15} models. We note a strong mismatch between observations and model predictions. The models for ages in the 10--25 Myr range, predict a substantial/complete depletion in the V$-$K$_s$ color range 3.8--4.5\,mag, whereas the Li gap is observed at colors that are redder by about 0.7\,mag. In fact, a better match is achieved if we arbitrarily shift the model colors by $\Delta$(V$-$K$_s$) = +0.7\,mag (see right panel of  Fig.\,\ref{LDB}).
Moreover, assuming that the fast rotators have rotation-unaffected Li abundances we have maintained their observed Li EW, whereas for the slower rotators we have 
increased the Li EW according to Equation (1).
 As it is shown in the right panel of Fig.\,\ref{LDB}, we find that, even applying this V$-$K$_s$ color shift and the rotation decorrelation, we can infer different ages by looking at different part of the observed Li depletion since the slope of the observed depletion is different than that predicted by the models.

\begin{figure}[!h]
\begin{minipage}{10cm}
\includegraphics[scale = 0.3, trim = 0 0 0 0, clip, angle=90]{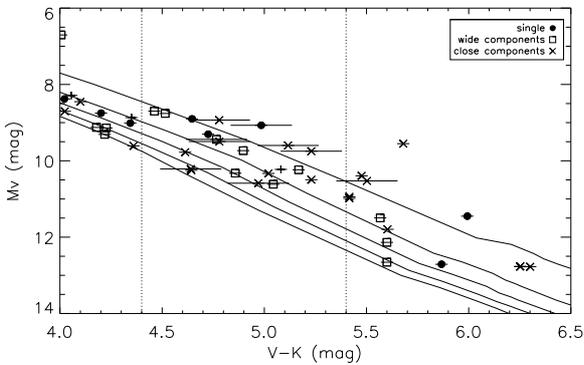}
\end{minipage}
\caption{\label{hr} Absolute magnitude versus V$-$K$_s$ color of members in the Li gap. Solid lines represent the isochrones corresponding to ages from 5 (top) to 25\,Myr (bottom).}
\end{figure}


We note that the same mismatch between observations and models is found when different colors (e.g., V$-$I, J$-$H, or J$-$K) \rm or the effective temperatures are used to make the comparison.
The colors of the stars in the Li depletion gap  4.4 $<$ V$-$K$_s$ $<$ 5.4\,mag have been accurately checked.
For a few stars (either single or resolved components in wide binaries)
the V$-$K$_s$ is measured for each component. For other stars (components of unresolved binaries) the V$-$K$_s$ is derived
from the spectral type of the primary component, which is the one with measured Li. 
In the first case the error on V$-$K$_s$ is $\pm$0.03\,mag, in the latter case
is $\pm$0.15\,mag.
In Fig.\,\ref{hr} we compare the absolute magnitudes of the members in the Li gap with isochrones for ages from 5 to 25 Myr taken from \citet{Baraffe15}. We note that within the gap (4.4 $<$V$-$K$_s$ $<$ 5.4\,mag) the stars seem to be  either more luminous or redder \rm than expected for the  most recent age estimation of 24\,Myr by \citet{Bell15}. \rm A better agreement is observed at bluer colors where stars are within the 10--25 Myr isochrones. If we correct the magnitudes of close components for binarity, we get a better agreement, but do not solve the overluminosity.
  This indicates that, in order to reproduce the data, the predicted \citet{Baraffe15} colours should be shifted redward.\\

\begin{figure*}
\begin{minipage}{20cm}
\includegraphics[scale = 0.3, trim = 0 0 0 0, clip, angle=90]{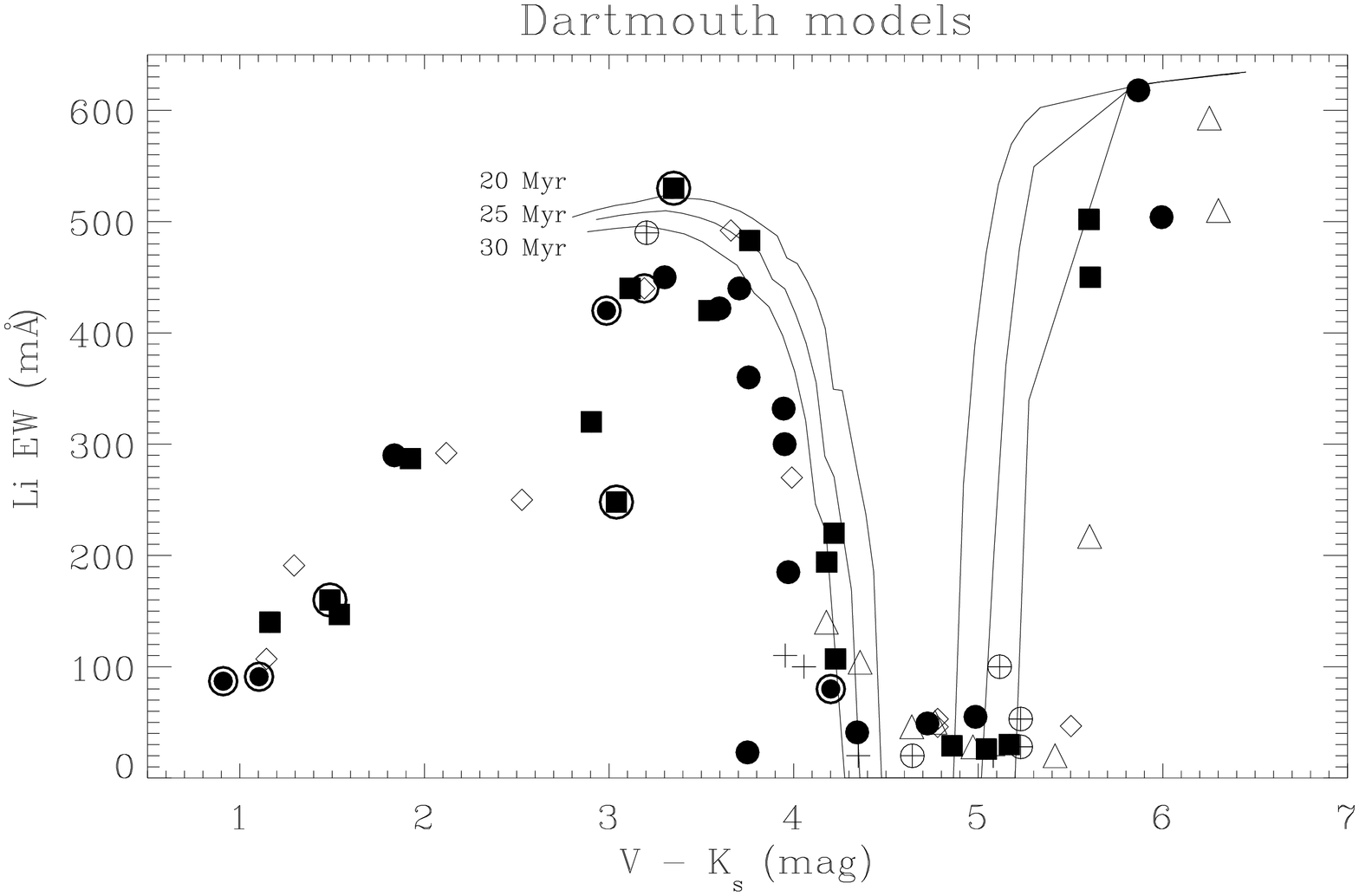}
\includegraphics[scale = 0.3, trim = 0 0 0 0, clip, angle=90]{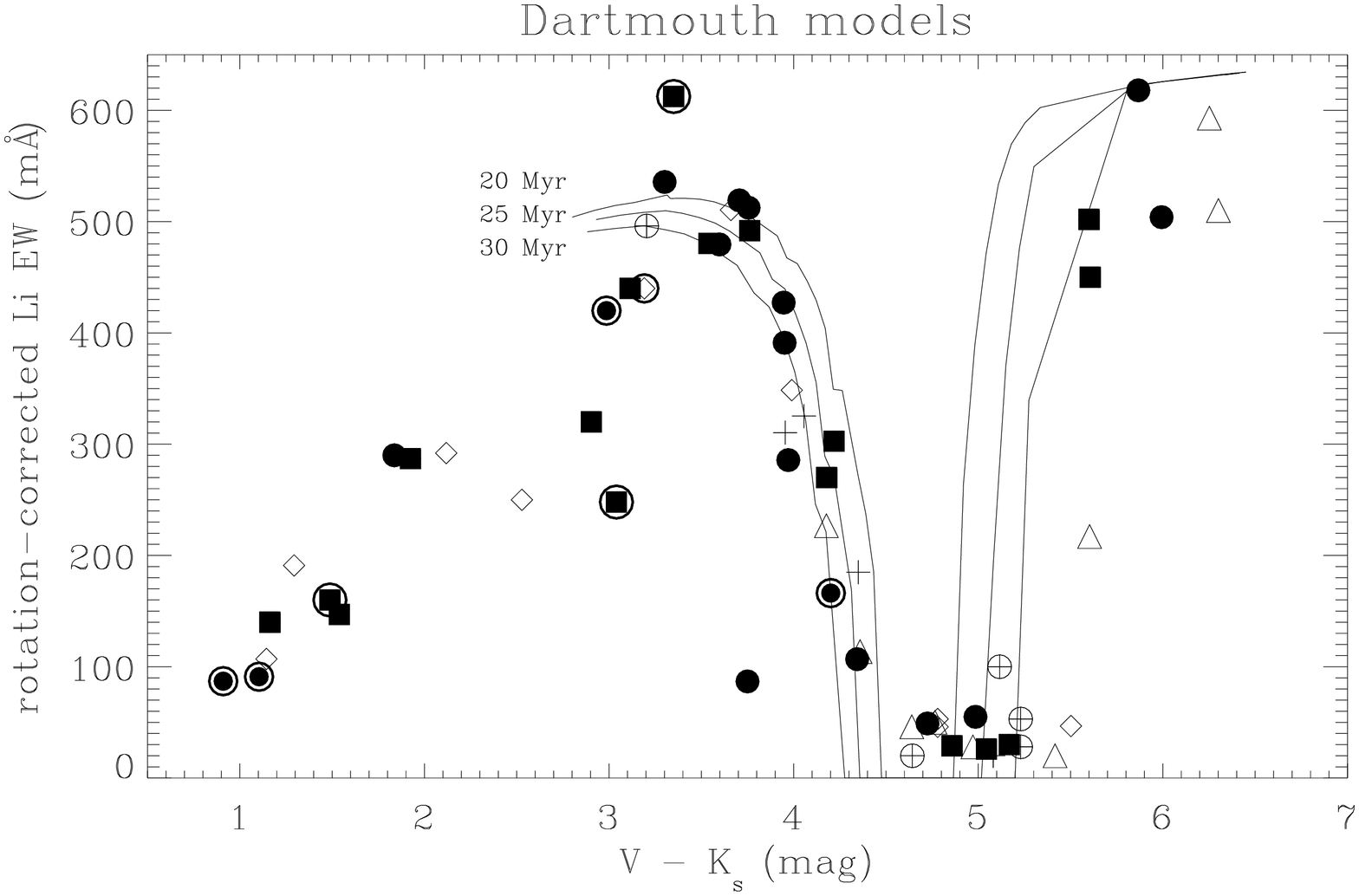}
\end{minipage}
\caption{\label{LDB_Dart} \it Left panel: \rm  distribution of Li EWs versus V$-$K$_s$ color for $\beta$ Pic members. Solid lines represent the Li EWs predicted by the Dartmouth models for ages in the range from 20 to 30 Myr. \it Right panel: \rm same as in the left panel, but Li EW decorrelated from rotation, assuming that fast rotators have rotation-unaffected Li EW.}
\end{figure*}

All stars in the gap have a prominent magnetic activity as it is inferred from photometric variability, X-ray emission and Near-UV/Far-UV excess. Therefore, models including the effect of magnetic fields should be more adequate to describe the Li depletion pattern. For this reason, we also made use of the Dartmouth stellar evolutionary models that incorporate the effects of magnetic fields \citep{Feiden16}; \citep{Mann16}. In these models, inhibition of convection by magnetic fields cools the  stellar  surface  temperature thereby slowing  the  contraction  rate  of  young  stars.  Stars  have  a  larger  radius  and  a higher luminosity at a given age, as a result

These models were computed for solar metallicity and for an equipartition magnetic field strength in the range 2500 $<$ Bf$_{eq}$ $<$ 3000\,G. Model effective temperatures were transformed into V$-$K$_s$ colors using the empirical T$_{\rm eff}$-V$-$K$_s$ relation from \citet{Pecaut13} valid for young 5--30\,Myr stars. Li abundances were transformed into Li EW as done in the case of the Baraffe et al. models.

In the left panel of Fig.\,\ref{LDB_Dart}, we plot the measured Li EW and compare this distribution with the Li EW derived from the Dartmouth et al. models. In this case, no arbitrary color shift is needed to match with the hot side of the Li gap. A better match is achieved in the right panel of Fig.\,\ref{LDB_Dart}, when the Li EW are decorrelated from rotation,  assuming again that the fast rotators have rotation-unaffected Li abundances.

\begin{figure}[!h]
\begin{minipage}{10cm}
\includegraphics[scale = 0.3, trim = 0 0 0 0, clip, angle=90]{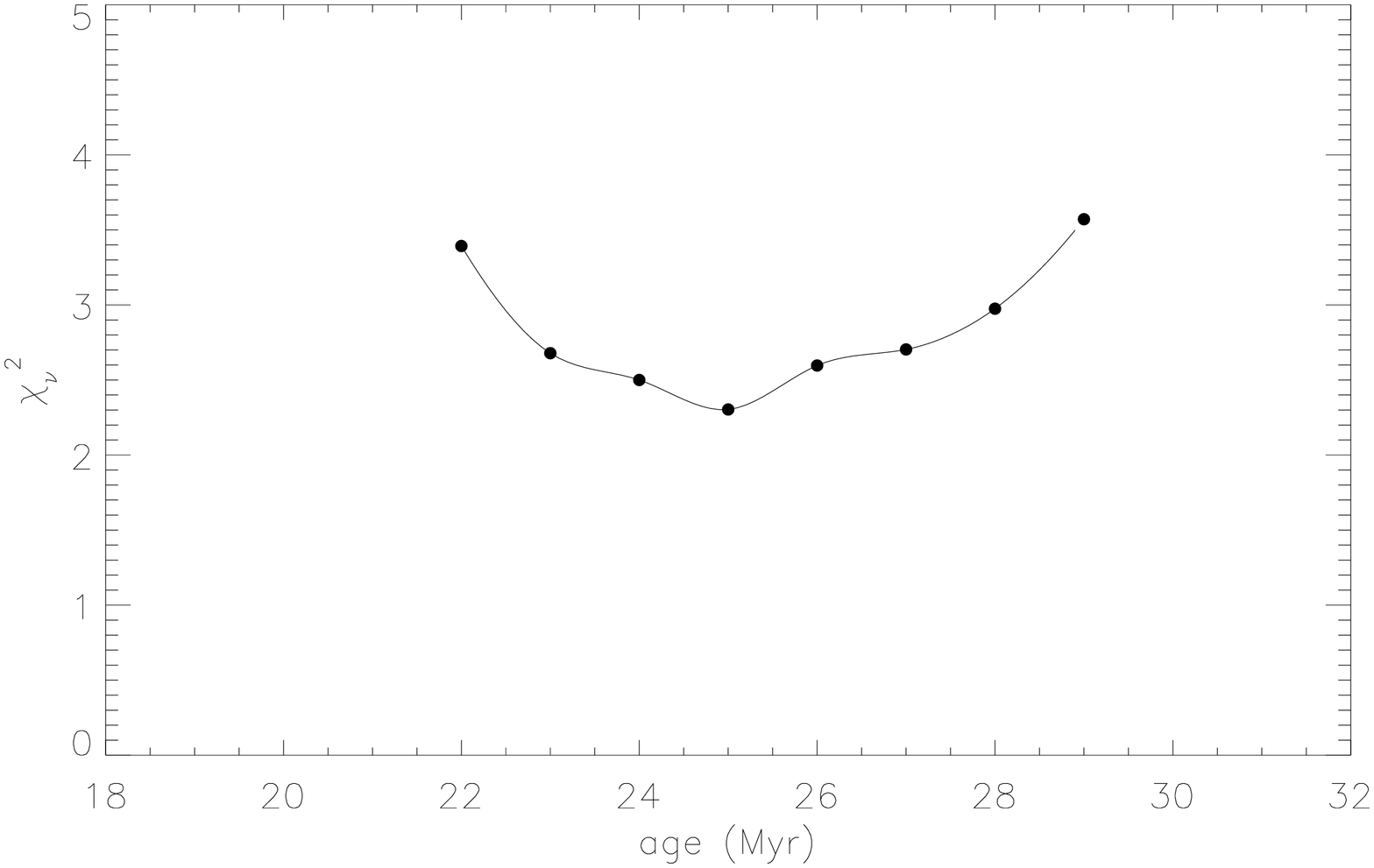}
\end{minipage}
\caption{\label{chisq} Reduced chi-squares versus age obtained from the residuals from the fit
to the hot boundary of the Li depletion gap.}
\end{figure}

The reduced chi-squares\footnote{To compute the  $\chi^{2}_{\nu}$ we used the uncertainty $\sigma = \sqrt{\sigma_0^2+\sigma_v^2}$ where $\sigma_0$ is the measurement error and  $\sigma_v$  is the average intrinsic variability. } computed from the fit to the hot boundary of the Li gap have their minimum (see 
Fig.\,\ref{chisq}) for an age of 25$\pm$3\,Myr. The uncertainty formally represents the age interval by which the $\chi^{2}_{\nu}$ increases by one unity from the minimum value. The relatively large values of the $\chi^{2}_{\nu}$ indicate either that the EW scatter arising from the intrinsic variability (rotational modulation) is larger than our estimate of 26\,m\AA\,\,or  the uncertainties associated to the measurements are in several case underestimated to some level.\\
Our age estimation is in good agreement with respect to the more recent age estimate of 23$\pm$4\,Myr and 24$\pm$3\,Myr by \citet{Mamajek14} and \citet{Bell15}, respectively, and significantly lower than the \citet{Macdonald10} estimate. \rm


\section{Conclusions}

We have retrieved from the literature the measured Li EW of 66  members of the young $\beta$ Pictoris association.
We have carried out a photometric monitoring of these members that allowed us to measure for the first time the rotation periods of 30 members. For other 16 members we retrieved the rotation periods from \citet{Messina10, Messina11}, and for the remaining members we used different sources in the literature.\\
We have explored the existence of a connection between rotation and  Li depletion. After removing the mass dependence of the Li EW, using linear fits to EW versus V$-$K$_s$, we found that for stars with 0.5 $\le$ V$-$K$_s$ $\le$ 3.4\,mag, roughly corresponding to  masses M $>$ 0.8\,M$_\odot$, no significant correlation is found between Li EW and rotation period. On the contrary, in the color range 3.4 $\le$ V$-$K$_s$ $\le$ 4.5\,mag, roughly corresponding to  masses 0.3 $<$ M $<$ 0.8\,M$_\odot$, we find a strong correlation between the Li EW and the 
rotation period, where fast rotators are much less Li-depleted than slow rotators. Finally, in the color range 5.4 $<$ V$-$K$_s$ $<$ 5.9\,mag, roughly corresponding to  masses M $\sim$ 0.1\,M$_\odot$, we have some hint for an inverted correlation where fast rotators are more depleted than slow rotators. However, this correlation is currently based on only 7 stars and the significance level is  high (80\%), but this is due to just one point that could be an outlier for whatever reason.\\
Interestingly, the dispersion in the 3.4 $\le$ V$-$K$_s$ $\le$ 4.5\,mag has a peak-to-peak amplitude that  amounts to 180\,m\AA\,\, which is about a factor of 2 larger than that measured in the same mass range in the 5-Myr \object{NGC\,2264} open cluster \citep{Bouvier16}, and about a factor 2 smaller than that measured in the 125-Myr \object{Pleiades} open cluster \citep{Soderblom93}. Therefore, we note that the effect of rotation on the Li depletion is also age dependent and increases with age.\\
A comparison with the \citet{Baraffe15} models, shows a mismatch of about 0.7\,mag  in the V$-$K$_s$ color \rm between the observed and the predicted color range where the Li gap falls. Models predict the Li gap at bluer colors than observed. 
 On the contrary, the Dartmouth models that incorporate the effects of magnetic fields provide a good match with the hot side of the Li depletion gap, although some mismatch on the cool side of the Li gap remains. Comparing the Dartmouth models with the hot side of the Li depletion gap, we infer an age of 25$\pm$3\,Myr, which is in very good agreement with the most recent age estimates for the $\beta$ Pictoris association. However, the relatively large values of the reduced chi-squares suggest that
either the intrinsic Li EW variability is still underestimated or some other factor, apart from rotation, plays a relevant role in producing the observed scatter among stars with similar mass.\\
\rm

{\it Acknowledgements}. Research on stellar activity at INAF- Catania Astrophysical Observatory is supported by MIUR  (Ministero dell'Istruzione, dell'Universit\'a e della Ricerca).  This research has made use of the Simbad database, operated at CDS (Strasbourg, France). SM thanks Jerome Bouvier for useful discussion and the anonymous Referee for useful comments that helped us to improve this paper.

\begin{table*}
\caption{Properties of the 66 members of the $\beta$ Pictoris association studied in this work}
\begin{minipage}{20cm}
\hspace{-1cm}
\scriptsize
\begin{tabular}{l l l l l l l c r r r r}
\hline
Target &   
  \multicolumn{3}{c}{RA } &
   \multicolumn{3}{c}{DEC} &
  Sp.T &
V\hspace{0.3cm} &
V$-$K$_s$ &
P \hspace{0.2cm} &
Li EW\\
 & \multicolumn{3}{c}{(hh, mm, ss)} & \multicolumn{3}{c}{($^{\circ}$, $^{\prime}$, $^{\prime\prime}$)} & & (mag) & (mag) & (d)\hspace{0.2cm}  &  (m\AA)\\
\hline
                      \object{HIP\,560}&   00&   06&  50.08&  $-$23&   06&  27.20&           F3V&    6.15  &  0.910  &  0.224  &   87.0 \\
                \object{TYC\,1186-0706-1}&   00&  23&  34.66&   +20&  14&  28.75&      K7.5V+M5&   10.96  &  3.623  &  7.900  &  338.0 \\
                       \object{GJ\,2006A}&   00&  27&  50.23&  $-$32&  33&   06.42&        M3.5Ve&   12.87  &  4.858  &  3.990  &   29.0 \\
                       \object{GJ\,2006B}&   00&  27&  50.35&  $-$32&  33&  23.86&        M3.5Ve&   13.16  &  5.044  &  4.910  &   26.0 \\
            \object{2MASS\,J01112542+1526214}A&   01&  11&  25.42&   +15&  26&  21.50&         M5+M6&   14.46  &  6.252  &  0.911  &  593.0 \\
             \object{2MASS\,J01351393-0712517}&   01&  35&  13.93&   $-$07&  12&  51.77&         M4.5V&   13.22  &  5.502  &  0.703  &   46.7 \\
               \object{TYC\,1208-0468-1}&   01&  37&  39.42&   +18&  35&  32.91&         K3+K5&    9.85  &  3.114  &  2.803  &  440.0 \\
                      \object{HIP\,10679}&   02&  17&  24.74&   +28&  44&  30.43&           G2V&    7.75  &  1.488  &  0.777  &  160.0 \\
                      \object{HIP\,10680}&   02&  17&  25.28&   +28&  44&  42.16&           F5V&    6.95  &  1.163  &  0.240  &  140.0 \\
                     \object{HIP\,11437}A&   02&  27&  29.25&   +30&  58&  24.60&            K4&   10.12  &  3.040  & 12.500  &  248.0 \\
                     \object{HIP\,11437}B&   02&  27&  28.05&   +30&  58&  40.53&            M1&   12.44  &  4.219  &  4.660  &  220.0 \\
                      \object{HIP\,12545}&   02&  41&  25.90&   +05&  59&  18.00&          K6Ve&   10.37  &  3.301  &  4.830  &  450.0 \\
                        \object{GJ3305}&   04&  37&  37.30&   $-$02&  29&  28.00&         M1+M?&   10.59  &  4.177  &  4.890  &  140.0 \\
             \object{2MASS\,J04435686+3723033}&   04&  43&  56.87&   +37&  23&   03.30&      M3Ve+M5?&   12.98  &  4.179  &  4.288  &  194.0 \\
                      \object{HIP\,23200}&   04&  59&  34.83&    +01&  47&   00.68&        M0.5Ve&   10.05  &  3.990  &  4.430  &  270.0 \\
                      \object{HIP\,23309}&   05&   00&  47.10&  $-$57&  15&  25.00&          M0Ve&   10.00  &  3.756  &  8.600  &  360.0 \\
                     \object{HIP\,23418}A&   05&   01&  58.80&    +09&  59&   00.00&           M3V&   11.45  &  4.780  &  1.220  &   53.0 \\
                    \object{BD\,-211074A}&   05&   06&  49.90&  $-$21&  35&   09.00&         M1.5V&   10.29  &  4.350  &  9.300  &   20.0 \\
                    \object{BD\,-211074B}&   05&   06&  49.90&  $-$21&  35&   09.00&         M2.5V&   11.67  &  4.643  &  5.400  &   20.0 \\
             \object{2MASS\,J05082729-2101444}&   05&  08&  27.30&  $-$21&   01&  44.40&         M5.6V&   14.70  &  5.867  &  0.280  &  618.0 \\
             \object{2MASS\,J05241914-1601153}&   05&  24&  19.15&  $-$16&   01&  15.30&       M4.5+M5&   13.50  &  5.603  &  0.401  &  217.0 \\
                      \object{HIP\,25486}&   05&  27&   04.76&  $-$11&  54&   03.47&           F7V&    6.22  &  1.294  &  0.966  &  191.0 \\
             \object{2MASS\,J05335981-0221325}&   05&  33&  59.81&   $-$02&  21&  32.50&         M2.9V&   12.42  &  4.725  &  7.250  &   49.0 \\
             \object{2MASS\,J06131330-2742054}&   06&  13&  13.31&  $-$27&  42&   05.50&         M3.V:&   12.09  &  5.230  & 16.9  &   28.0 \\
                      \object{HIP\,29964}&   06&  18&  28.20&  $-$72&   02&  41.00&          K4Ve&    9.80  &  2.986  &  2.670  &  420.0 \\
                         \object{TWA\,22}&  10&  17&  26.89&  $-$53&  54&  26.50&            M5&   13.99  &  6.301  &  0.830  &  510.0 \\
                      \object{HIP\,76629}&  15&  38&  57.50&  $-$57&  42&  27.00&           K0V&    7.97  &  2.118  &  4.270  &  292.0 \\
             \object{2MASS\,J16430128-1754274}&  16&  43&   01.29&  $-$17&  54&  27.50&          M0.6&   12.50  &  3.951  &  5.140  &  300.0 \\
                      \object{HIP\,84586}&  17&  17&  25.50&  $-$66&  57&   04.00&     G5IV+K5IV&    7.23  &  2.528  &  1.680  &  250.0 \\
                     \object{HD\,155555C}&  17&  17&  31.29&  $-$66&  57&   05.49&        M3.5Ve&   12.71  &  5.081  &  4.430  &   20.0 \\
                  \object{TYC\,872822621}&  17&  29&  55.10&  $-$54&  15&  49.00&           K1V&    9.55  &  2.186  &  1.830  &  360.0 \\
                \object{GSC\,08350-01924}&  17&  29&  20.67&  $-$50&  14&  53.00&           M3V&   12.86  &  4.766  &  1.982  &   50.0 \\
                      \object{V4046\,Sgr}&  18&  14&  10.50&  $-$32&  47&  33.00&         K5+K7&   10.44  &  3.191  &  2.420  &  440.0 \\
             \object{2MASS\,J18151564-4927472}&  18&  15&  15.64&  $-$49&  27&  47.20&           M3V&   12.86  &  4.780  &  0.447  &   46.0 \\
                      \object{HIP\,89829}&  18&  19&  52.20&  $-$29&  16&  33.00&           G1V&    8.89  &  1.837  &  0.571  &  290.0 \\
            \object{2MASS\,J18202275-1011131}A&  18&  20&  22.74&  $-$10&  11&  13.62&     K5Ve+K7Ve&   10.63  &  3.350  &  4.650  &  530.0 \\
                  \object{TYC\,907724891}&  18&  45&  37.02&  $-$64&  51&  46.14&          K5Ve&    9.30  &  3.204  &  0.345  &  490.0 \\
                  \object{TYC\,907307621}&  18&  46&  52.60&  $-$62&  10&  36.00&          M1Ve&   11.80  &  3.946  &  5.370  &  332.0 \\
                      \object{HD\,173167}&  18&  48&   06.36&  $-$62&  13&  47.02&           F5V&    7.28  &  1.144  &  0.250  &  107.0 \\
                  \object{TYC\,740800541}&  18&  50&  44.50&  $-$31&  47&  47.00&          K8Ve&   11.20  &  3.660  &  1.075  &  492.0 \\
                      \object{HIP\,92680}&  18&  53&   05.90&  $-$50&  10&  50.00&          K8Ve&    8.29  &  1.924  &  0.944  &  287.0 \\
                  \object{TYC\,687210111}&  18&  58&   04.20&  $-$29&  53&   05.00&          M0Ve&   11.78  &  3.762  &  0.503  &  483.0 \\
            \object{2MASS\, J19102820-2319486}&  19&  10&  28.21&  $-$23&  19&  48.60&           M4V&   13.20  &  4.985  &  3.640  &   55.0 \\
                  \object{TYC\,687801951}&  19&  11&  44.70&  $-$26&   04&   09.00&          K4Ve&   10.27  &  2.904  &  5.650  &  320.0 \\
             \object{2MASS\,J19233820-4606316}&  19&  23&  38.20&  $-$46&   06&  31.60&           M0V&   11.87  &  3.598  &  3.242  &  422.0 \\
                  \object{TYC\,744311021}&  19&  56&   04.37&  $-$32&   07&  37.71&         M0.0V&   11.80  &  3.954  & 11.300  &  110.0 \\
          \object{2MASS\, J19560294-3207186}AB&  19&  56&   02.94&  $-$32&  07&  18.70&           M4V&   13.23  &  5.116  &  1.569  &  100.0 \\
             \object{2MASS\,J20013718-3313139}&  20&   01&  37.18&  $-$33&  13&  14.01&            M1&   12.25  &  4.056  & 12.700  &  100.0 \\
             \object{2MASS\,J20055640-3216591}&  20&   05&  56.41&  $-$32&  16&  59.15&           M2:&   11.96  &  4.022  &  8.368  &  140.0 \\
                      \object{HD\,191089}&  20&   09&   05.21&  $-$26&  13&  26.52&           F5V&    7.18  &  1.104  &  0.488  &   91.0 \\
          \object{2MASS\, J20100002-2801410}AB&  20&  10&   00.03&  $-$28&   01&  41.10&     M2.5+M3.5&   12.80  &  4.640  &  0.470  &   46.0 \\
             \object{2MASS\,J20333759-2556521}&  20&  33&  37.59&  $-$25&  56&  52.20&          M4.5&   14.87  &  5.993  &  0.710  &  504.0 \\
                    \object{HIP\,102141}A&  20&  41&  51.20&  $-$32&  26&   07.00&          M4Ve&   10.36  &  5.416  &  1.191  &   20.0 \\
                    \object{HIP\,102141}B&  20&  41&  51.10&  $-$32&  26&  10.00&          M4Ve&   10.36  &  5.416  &  0.781  &   20.0 \\
             \object{2MASS\,J20434114-2433534}&  20&  43&  41.14&  $-$24&  33&  53.19&     M3.7+M4.1&   12.73  &  4.971  &  1.610  &   28.0 \\
                     \object{HIP\,102409}&  20&  45&   09.50&  $-$31&  20&  27.00&          M1Ve&    8.73  &  4.201  &  4.860  &   80.0 \\
                     \object{HIP\,103311}&  20&  55&  47.67&  $-$17&   06&  51.04&           F8V&    7.35  &  1.539  &  0.356  &  147.0 \\
                  \object{TYC\,634902001}&  20&  56&   02.70&  $-$17&  10&  54.00&       K6Ve+M2&   10.62  &  3.541  &  3.410  &  420.0 \\
             \object{2MASS\,J21100535-1919573}&  21&  10&   05.36&  $-$19&  19&  57.40&           M2V&   11.54  &  4.344  &  3.710  &   41.0 \\
            \object{2MASS\, J21103147-2710578}&  21&  10&  31.48&  $-$27&  10&  57.80&         M4.5V&   14.90  &  5.600  &  0.650  &  502.0 \\
                 \object{TYC\,9486-927-1}&  21&  25&  27.49&  $-$81&  38&  27.68&           M2V&   11.70  &  4.360  &  0.542  &  104.0 \\
		\object{TYC\,221113091} & 22 & 00 & 41.59  & +27 & 15  & 13.60 &  M0V & 11.39 & 3.666 & 1.109 &  40.0 \\
                  \object{TYC\,934004371}&  22&  42&  48.90&  $-$71&  42&  21.00&          K7Ve&   10.60  &  3.706  &  4.460  &  440.0 \\
                     \object{HIP\,112312}&  22&  44&  58.00&  $-$33&  15&   02.00&          M4Ve&   12.10  &  5.168  &  2.370  &   30.0 \\
                         \object{TX Psa}&  22&  45&   00.05&  $-$33&  15&  25.80&        M4.5Ve&   13.36  &  5.567  &  1.080  &  450.0 \\
                  \object{TYC\,583206661}&  23&  32&  30.90&  $-$12&  15&  52.00&          M0Ve&   10.54  &  3.971  &  5.680  &  185.0 \\
                  \hline
\end{tabular}
\end{minipage}
\end{table*}

\bibliographystyle{aa} 
\bibliography{mybib} 
\end{document}